\documentclass[a4paper,fleqn]{cas-dc}

\usepackage[authoryear,longnamesfirst]{natbib}

\def\tsc#1{\csdef{#1}{\textsc{\lowercase{#1}}\xspace}}
\tsc{WGM}
\tsc{QE}

\begin{document}
\let\WriteBookmarks\relax
\def\floatpagepagefraction{1}
\def\textpagefraction{.001}

\shorttitle{Explainable Deep Learning Algorithm for Distinguishing Incomplete Kawasaki Disease}    

\shortauthors{H. Lee \textit{et al.}}  

\title [mode = title]{Explainable Deep Learning Algorithm for Distinguishing Incomplete Kawasaki Disease by Coronary Artery Lesions on Echocardiographic Imaging}

\author[1]{Haeyun Lee}[type=author,
      orcid=0000-0002-7572-1705]

\author[1]{Yongsoon Eun}[type=author,
      orcid=0000-0002-2304-7106]

\author[1]{Jae Youn Hwang}[type=author,
      orcid=0000-0003-4659-6009]
      
\cormark[1]
\ead{jyhwang@dgist.ac.kr}

\author[2]{Lucy Youngmin Eun}[type=author,
      orcid=0000-0002-4577-3168]
      
\cormark[1]
\ead{lucyeun@yuhs.ac}

\affiliation[1]{organization={Department of Information and Communication Engineering, Daegu Gyeongbuk Institute of Science \& Technology},
            city={Daegu},
            postcode={42988}, 
            country={Korea}}
            
\affiliation[2]{organization={Division of Pediatric Cardiology, Department of Pediatrics, Yonsei University College of Medicine},
            city={Seoul},
            postcode={03722}, 
            country={Korea}}

\cortext[1]{Corresponding author}



\begin{abstract}
Background and Objective: Incomplete Kawasaki disease (KD) has often been misdiagnosed due to a lack of the clinical manifestations of classic KD. However, it is associated with a markedly higher prevalence of coronary artery lesions. Identifying coronary artery lesions by echocardiography is important for the timely diagnosis of and favorable outcomes in KD. Moreover, similar to KD, coronavirus disease 2019, currently causing a worldwide pandemic, also manifests with fever; therefore, it is crucial at this moment that KD should be distinguished clearly among the febrile diseases in children. In this study, we aimed to validate a deep learning algorithm for classification of KD and other acute febrile diseases. 

Methods: We obtained coronary artery images by echocardiography of children (n = 88 for KD; n = 65 for pneumonia). We trained six deep learning networks (VGG19, Xception, ResNet50, ResNext50, SE-ResNet50, and SE-ResNext50) using the collected data.

Results: SE-ResNext50 showed the best performance in terms of accuracy, specificity, and precision in the classification. SE-ResNext50 offered a precision of 76.35\%, a sensitivity of 82.64\%, and a specificity of 58.12\%.

Conclusions: The results of our study suggested that deep learning algorithms have similar performance to an experienced cardiologist in detecting coronary artery lesions to facilitate the diagnosis of KD.

\end{abstract}




\maketitle

\section{Introduction}\label{intro}
Kawasaki disease (KD) is the most common acquired heart disease in childhood. It was first described by Dr. Kawasaki in 1967~\cite{kuo2017preventing}. 
KD mostly occurs in children aged less than 5 years and has a high prevalence in countries of Northeast Asia, particularly Japan, South Korea, and Taiwan~\cite{makino2015descriptive, kim2017epidemiology}.

The main symptoms of KD are unexplained high fever, diffuse erythematous polymorphous rash, bilateral conjunctival injection, cervical lymphadenopathy, oral mucosal changes, and extremity changes, sometimes with perineal desquamation, or reactivation of the bacillus Calmette–Guérin injection site~\cite{mccrindle2017american, dietz2017dissecting, newburger2016kawasaki, singh2018diagnosis}. 
While the widely used diagnostic criteria for KD are useful, incomplete KD in infants or children aged 10 years or older can often be problematic, causing misdiagnosis due to a lack of manifestation of the full clinical criteria of classic KD. 
Nevertheless, this disease has a much higher prevalence of coronary artery lesions~\cite{mccrindle2017american, singh2018diagnosis}.

Given the difficulty in diagnosing incomplete or “atypical” KD by clinical features alone, identifying coronary artery findings by echocardiography, along with evaluation of various biomarkers by blood tests, becomes more significant for ensuring an appropriate diagnosis~\cite{mccrindle2017american}. 
Even though the choice of treatment with a high dose of intravenous immunoglobulin infusion decreases the risk of coronary artery complications, about 5\% of treated children and 15–25\% of untreated children have a risk of coronary artery aneurysms or ectasia. 
Certainly, one of the fatal complications of untreated KD is coronary artery aneurysm. Accordingly, the role of echocardiography in recognizing coronary artery lesions is substantial to ensure timely diagnosis and favorable outcome~\cite{na2019utilization}.

For the proper diagnosis of incomplete KD, the expert pediatric cardiologists need to perform echocardiography to investigate the patient’s coronary arteries. 
The most important therapeutic goal of KD is the prevention of coronary artery aneurysm formation. When an aneurysm is noticed, it is critical to prevent development of a giant aneurysm or formation of a thrombus~\cite{rowley1988prevention}.

Unfortunately, without an experienced pediatric cardiologist and a KD expert, it is challenging to diagnose incomplete KD because the fever patterns of many acute febrile diseases in children appear similar to KD, particularly the initial high grade of fever. 
In 2020, reports of severely ill pediatric cases have shown that KD and coronavirus disease 2019 (COVID-19) presented similar symptoms~\cite{jones2020covid, viner2020kawasaki}.
Moreover, the incidence of KD suddenly increased in Europe and the USA during the COVID-19 pandemic. 
In addition, KD-like multi-systemic inflammatory syndrome in children affected many children in Europe and in the USA. This is of particular concern, as it can result in missed or delayed KD diagnosis and treatment~\cite{guan2020clinical}.

Several studies on computer-aided diagnosis based on deep learning algorithms have shown good performance. 
Deep learning-based approaches for the diagnosis of cancer or lesions have been shown to perform well and have already surpassed human doctors in some categories~\cite{esteva2017dermatologist}. 
Additionally, deep learning algorithms have shown better performance in the medical vision field than conventional methods~\cite{liu2019deep}. 
Several deep learning algorithms have been proposed to diagnose various diseases, such as breast cancer~\cite{lee2020channel}, liver cancer~\cite{wu2014deep}, and thyroid nodules~\cite{ma2017pre} on ultrasound images. 
These medical deep learning algorithms have been proposed for computer-aided enhancement of diagnostic performance. However, deep learning algorithms have not yet been applied to KD diagnosis.

The purpose of this study was to assess whether explainable deep learning algorithms could be used to identify coronary artery lesions on echocardiographic images for the timely diagnosis of KD. 
We also evaluated the performance of these algorithms in distinguishing KD from another acute febrile disease, pneumonia.

\section{Methods}\label{methods}
\subsection{Data Acquisition}

To investigate whether it would be possible to distinguish between KD and another similar acute febrile disease, we selected pneumonia as an alternative representative acute febrile disease. 
Pneumonia is one of the most common febrile diseases in children.

For this study, echocardiographic imaging data from January 2016 to August 2019 were acquired from Yonsei University Gangnam Severance Hospital.
Giant coronary aneurysm cases were excluded from this study. 
Echocardiographic images of 88 children with incomplete KD and 59 children with pneumonia (147 in total) were acquired and labeled as KD and non-KD by an experienced cardiologist. 
2D echocardiographic coronary artery short axis view images were obtained for the appropriate diagnosis when the children initially presented with high grade fever. We cropped the echocardiographic images to $512 \times 512$ pixels.

\begin{figure*}[t]
    \centering
    \includegraphics[width=1\linewidth]{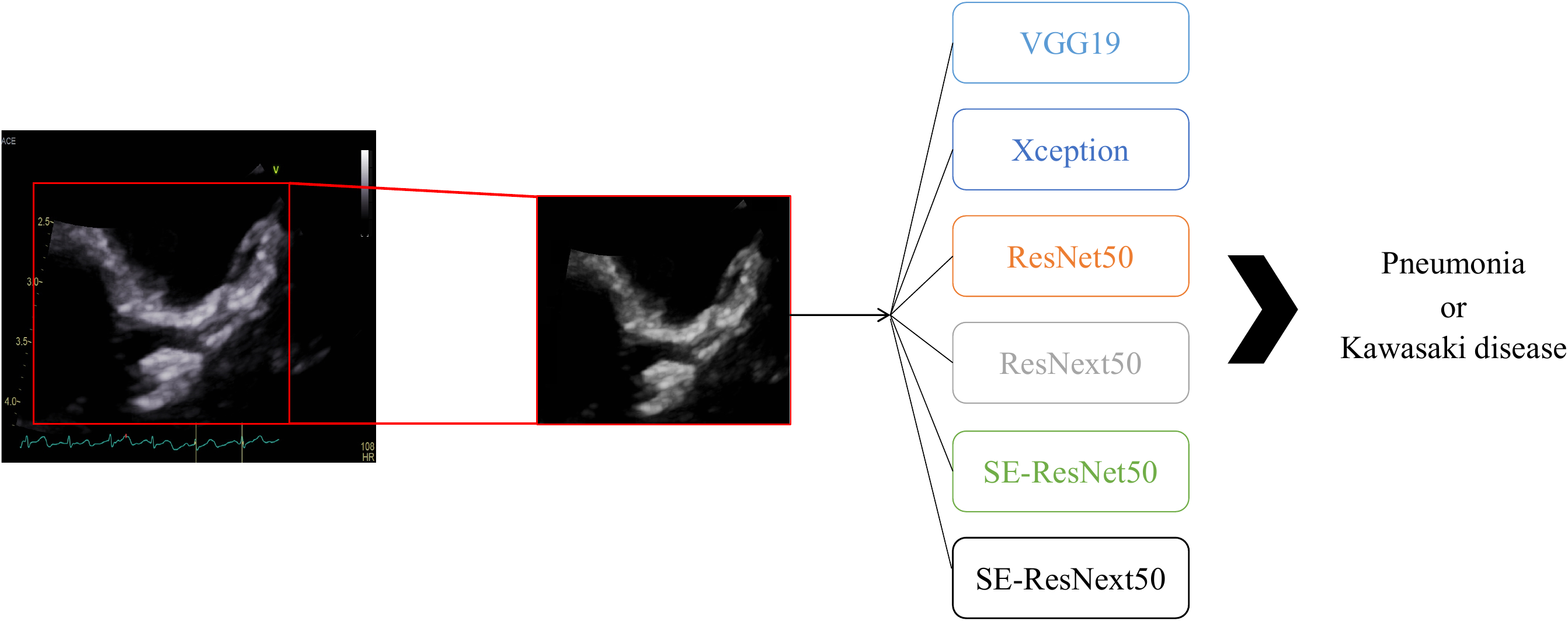}
    \caption{Echocardiography analysis process using deep learning algorithms in this study.}
    \label{fig:network_architecture}
\end{figure*}

\subsection{Deep Learning Algorithm}

In this study, to distinguish incomplete KD from non-KD using echocardiographic images, we applied six deep learning architectures: VGG-19~\cite{simonyan2014very}, Xception~\cite{chollet2017xception}, ResNet-50~\cite{he2016deep}, ResNext-50~\cite{xie2017aggregated}, SE-ResNet-50, and SE-ResNext-50~\cite{hu2018squeeze}.

VGG~\cite{simonyan2014very} is the most basic network with a simple structure for classification and good performance. 
Therefore, it is still widely used as a comparison architecture. 
Xception~\cite{chollet2017xception} is a linear stack of depth-wise separable convolutions with residual connections and shows higher performance than the baseline Inception architecture~\cite{szegedy2016rethinking}. 
ResNet~\cite{he2016deep} is a deep learning network which is stacked more deeply using a skip connection and is being used in various fields. 
ResNext~\cite{xie2017aggregated} has shown higher performance while reducing the computation cost compared to the existing ResNet by using a group convolution. 
SE-ResNet and SE-ResNext~\cite{hu2018squeeze} are networks in which squeeze-and-excitation blocks are added to ResNet and ResNext, respectively~\cite{russakovsky2015imagenet}. 

Since the data in our study were limited to training the models, only networks with 50 or fewer convolution layers were used in this experiment.
Thus, we used VGG-19, Xception, ResNet-50, ResNext-50, SE-ResNet50, and SE-ResNext50 in this experiment. We then evaluated the capability of the deep learning algorithms to distinguish between KD and non-KD.

For the training, we used a stochastic optimization method (ADAM)~\cite{kingma2014adam} with parameter $\beta_1=0.9$, $\beta_2=0.999$, and $\epsilon=10^{-8}$. 
The initial learning rate was $1e^{-3}$, and it decreased by 1/10 every 30 epochs. 
We trained each network for a total of 120 epochs. 
Training batch sizes consisted of 32 patches. 
We used a binary cross-entropy loss function to train each network. 
We used the pretrained weights of each network on ImageNet to achieve better performance~\cite{russakovsky2015imagenet}. 

The deep learning framework used for training and testing the deep learning algorithms was PyTorch~\cite{paszke2017automatic}. 
We trained and tested all networks using a 2-GHz Intel Xeon E5-2620 processor and an NVIDIA TITAN RTX graphics card (24 GB).

\begin{figure*}[t]
    \centering
    \includegraphics[width=0.8\linewidth]{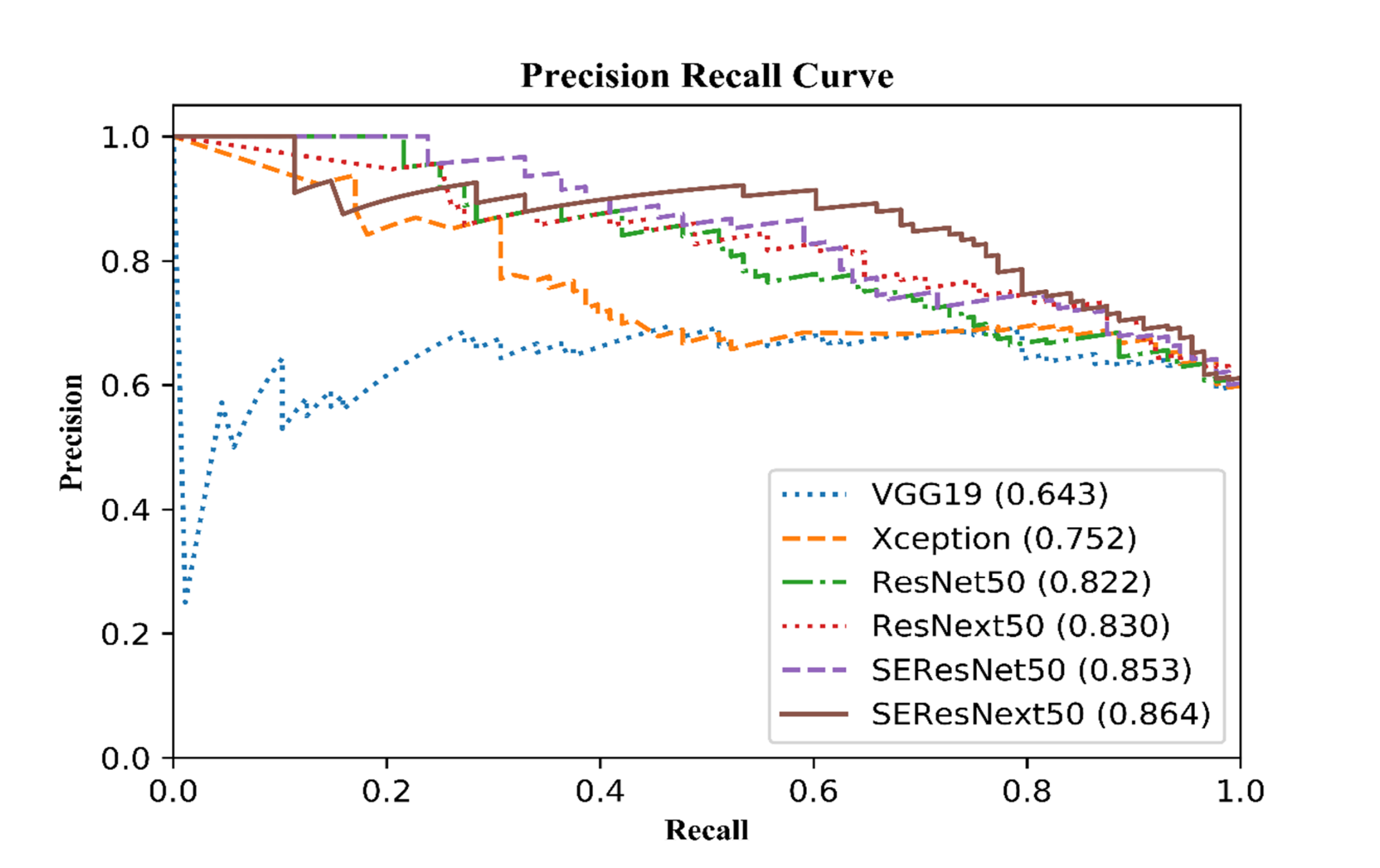}
    \caption{The precision-recall curves of each deep learning algorithm for KD. Areas under the precision-recall curve of VGG19, Xception, ResNet50, ResNext50, SE-ResNet50, and SE-ResNext50 were 0.643, 0.752, 0.822, 0.830, 0.853, and 0.864, respectively.}
    \label{fig:pr_curve}
\end{figure*}

\subsection{Class Activation Map}

To explain deep learning algorithms, we used the class activation map (CAM)~\cite{zhou2016learning}. 
Previously, it was not possible to know which salient parts of a medical image would be highlighted by a deep learning algorithm for classification. 
The CAM has been proposed to solve this issue and to be able to explain deep learning algorithms. 
Most deep learning algorithms use a fully connected layer to classify the values obtained by applying a global average pooling (GAP) to the feature maps from the last convolution layer. 
A linear transform with a class number of filters is then applied to the weight obtained through GAP.

Here, to obtain the CAM, the weight of the linear transform for each class was multiplied by the feature map obtained from the last convolutional layer. 
The CAM at class $c$, $M_c(x,y)$, can be calculated as follows:

\begin{equation}
\begin{aligned}
  M_c(x,y)& = \sum_k w_k^c F_k \\
  & = \sum_k w_k^c \sum_{x,y} f_k(x,y) \\
  & = \sum_{x,y} \sum_k w_k^c f_k(x,y) \\
\end{aligned}
\end{equation}
where $f_k(x,y)$ is a feature map from the last convolution layer for a unit k, $w_k^c$ is the weight of linear transformation corresponding to class $c$ for the unit $k$, and $x$ and $y$ are the spatial information of $f_k$ and $M_c$, respectively. 
The class activation map, $M_c(x,y)$, indicates a class-specific highlight map at a spatial grid $(x,y)$. 
Therefore, through the CAM, it is possible to understand which parts of the image are considered when the deep learning algorithm proceeds with classification. 
Hence, the CAM is an excellent tool for analyzing medical image deep learning algorithms~\cite{ma2020ms, qiao2019deep}, as in this study.

\begin{table*}[t]
    \centering
    \caption{Diagnostic performance of deep learning algorithms. The best performance is in \textbf{bold}, and the second best performance is in \underline{underlined}.}
    \begin{tabular}{|c|c|c|c|c|c|c|}
        \hline
        Networks & VGG19 & Xception & ResNet50 & ResNext50 & SE-ResNet50 & SE-ResNext50 \\
        \hline\hline
        Accuracy & 64.33 & 67.50 & 67.14 & \underline{71.60} & 70.42 & \textbf{72.88} \\
        \hline
        F1 score & 70.72 & 74.74 & 71.68 & \underline{77.17} & 75.56 & \textbf{78.26} \\
        \hline
        Sensitivity & 72.64 & \underline{80.70} & 70.69 & 77.13 & 77.64 & \textbf{82.64} \\
        \hline
        Specificity & 51.33 & 47.67 & 61.67 & 57.67 & \textbf{59.67} & \underline{58.12} \\
        \hline
        Precision (PPV) & 73.20 & 70.78 & 73.99 & \underline{75.10} & 74.61 & \textbf{76.35} \\
        \hline
        NPV & 55.56 & 62.22 & 58.06 & \underline{66.67} & 63.64 & \textbf{68.63} \\
        \hline
    \end{tabular}
    \label{tbl:quantitative_comparison}
\end{table*}

\subsection{Ethics Statement}

This study was approved by the Yonsei University College of Medicine Institutional Review Board and the Research Ethics Committee of Severance Hospital (study approval number: 2020-1127-001). 
All research was performed in accordance with relevant guidelines and regulations. The requirement for written informed consent was waived by the Institutional Review Board due to the retrospective study design.

\section{Results}\label{results}

\begin{figure*}[t]
    \centering
    \includegraphics[width=0.8\linewidth]{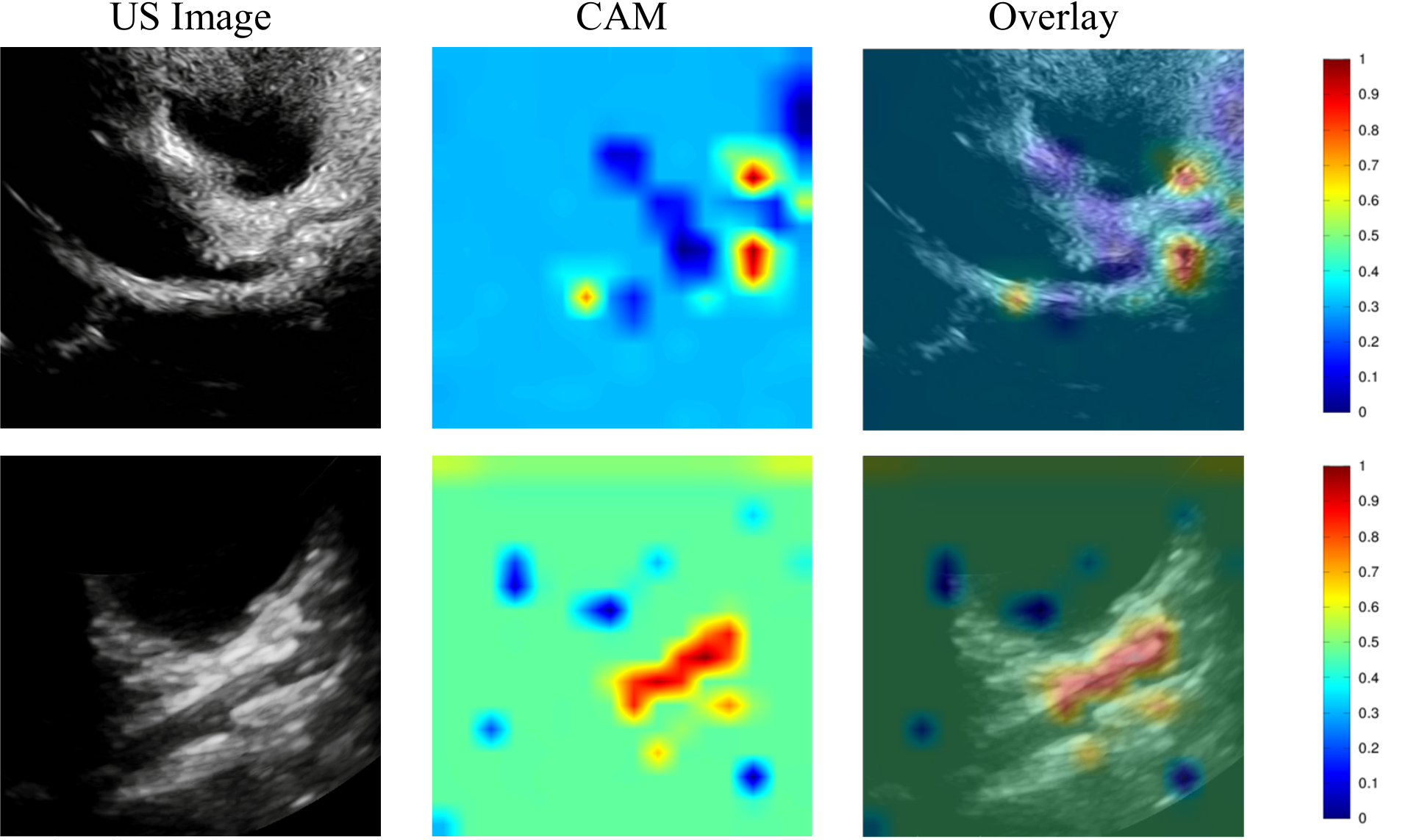}
    \caption{Illustration of a class activation map of SE-ResNext50 for non-KD: the first column shows the echocardiographic images of non-KD, the second column shows the CAMs, and the third column shows the overlay of the echocardiographic image and the CAM.}
    \label{fig:CAM1}
\end{figure*}

Since we conducted the experiments using 10-fold cross validation, there were 10 outcomes.
Each subset had 137 or 138 echocardiographic images for training and 16 or 15 echocardiographic images for testing. 
The test images in each subset included 8 or 9 echocardiographic images labeled as KD and 6 or 7 echocardiographic images labeled as non-KD (pneumonia) by an echocardiographic specialist. 
Table 1 shows the diagnostic performance of each network trained with each subset for the test dataset. 
The accuracies of VGG19, Xception, ResNet50, ResNext50, SE-ResNet50, and SE-ResNext50 were 64.33\%, 67.50\%, 58.58\%, 71.60\%, 63.72\%, and 72.88\% for the classification of KD and non-KD, respectively. 
The F1 score of VGG19, Xception, ResNet50, ResNext50, SE-ResNet50, and SE-ResNext50 were 70.72\%, 74.74\%, 71.68\%, 77.17\%, 75.56\%, and 78.26\% for the classification of KD and non-KD, respectively.
In contrast, the sensitivities of VGG19, Xception, ResNet50, ResNext50, SE-ResNet50, and SE-ResNext50 were 72.64\%, 80.70\%, 82.22\%, 77.13\%, 87.64\%, and 82.64\%, respectively. 
The specificities of VGG19, Xception, ResNet50, ResNext50, SE-ResNet50, and SE-ResNext50 were 51.33\%, 47.67\%, 23.33\%, 57.67\%, 28.00\%, and 58.12\%, respectively. 
The precision (positive predictive value; PPV) of VGG19, Xception, ResNet50, ResNext50, SE-ResNet50, and SE-ResNext50 were 73.20\%, 70.78\%, 62.36\%, 75.10\%, 65.18\%, and 76.35\%, respectively. 
The negative predictive value (NPV) of VGG19, Xception, ResNet50, ResNext50, SE-ResNet50, and SE-ResNext50 were 55.56\%, 62.22\%, 58.06\%, 66.67\%, 63.64\%, and 68.63\%, respectively.
In these results, SE-ResNext50 showed the best performance in terms of accuracy, F1 score, sensivitivity, precision, and NPV for the distinction of KD and non-KD.
It identified 112 true-positive diagnoses among 153 images. 
Figure 2 shows the precision-recall curve of the deep learning algorithms used for the classification between KD and pneumonia. 
The areas under the precision-recall curve (AUPRC) of VGG19, Xception, ResNet50, ResNext50, SE-ResNet50, and SE-ResNext50 were 0.643, 0.752, 0.822, 0.830, 0.853, and 0.864, respectively.

\begin{figure*}[t]
    \centering
    \includegraphics[width=0.8\linewidth]{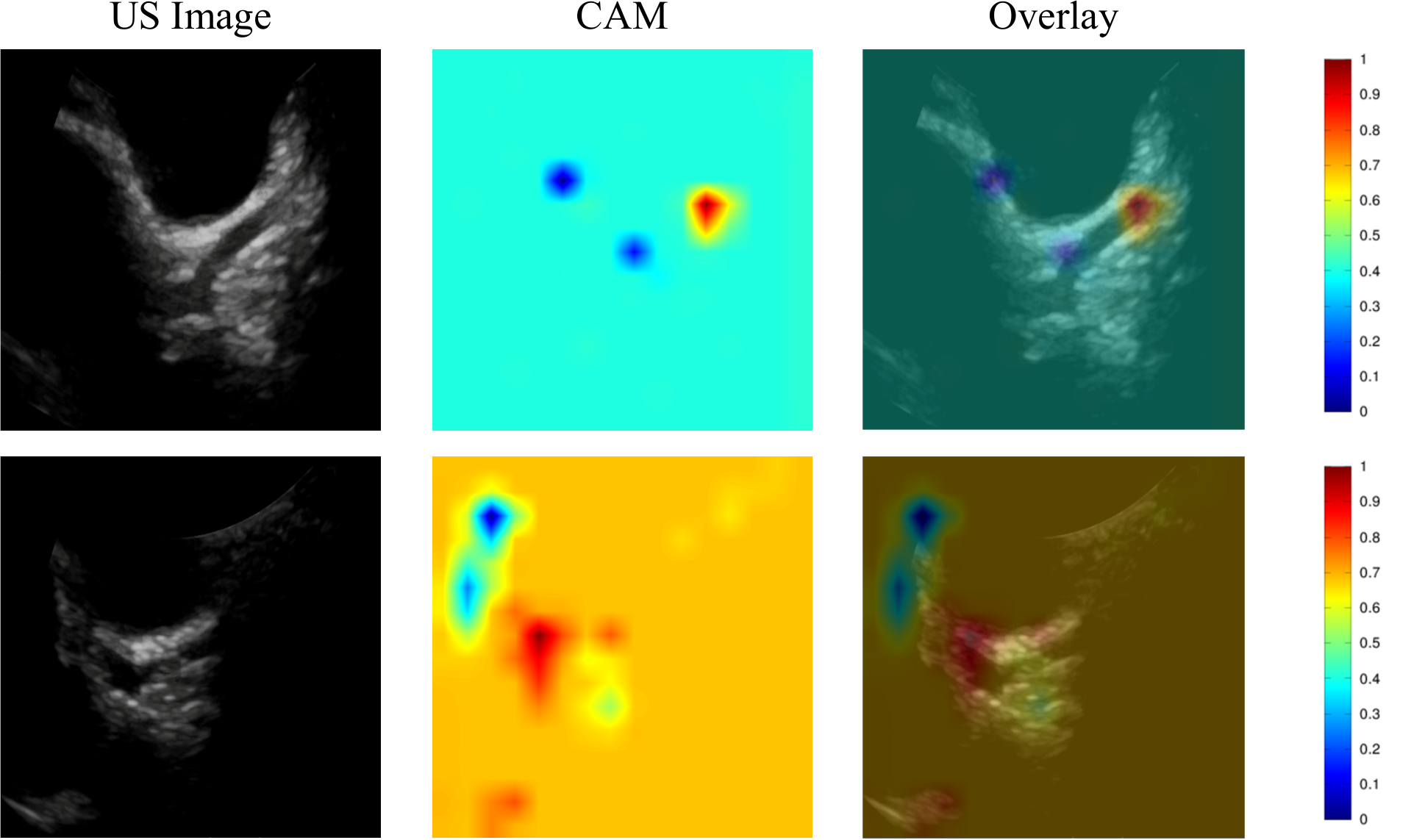}
    \caption{Illustration of a class activation map of SE-ResNext50 for KD: the first column shows the echocardiographic images of KD, the second column shows the CAMs, the third column shows the overlays of the echocardiographic images and the CAM.}
    \label{fig:CAM2}
\end{figure*}

For non-KD, the echocardiographic images of pneumonia and its corresponding CAM images are illustrated in Figure 3. 
These images were correctly identified as non-KD by SE-ResNext50. The images in the first row represent the pneumonia image correctly recognized as non-KD by SE-ResNext50, whereas the second row shows the pneumonia image incorrectly classified as KD by SE-ResNext50.

For incomplete KD, the echocardiographic image of KD and its corresponding CAM image are demonstrated in Figure 4. 
The images in the first row were correctly identified as KD by SE-ResNext50, whereas the images in the second row were incorrectly classified as non-KD by SE-ResNext50. 
The thicker red regions indicate the parts of images focused on by the deep learning algorithm during the process of classification as KD and non-KD.

\section{Discussion}\label{disc}
The goal of this study was to investigate the potential of explainable deep learning algorithms to identify and differentiate KD from acute febrile diseases.
We therefore selected several well-known deep learning algorithms (VGG19, Xception, ResNet50, ResNext50, SE-ResNet50, and SE-ResNext50) to distinguish incomplete KD from other acute febrile diseases. 
We selected pneumonia as a representative of other acute febrile diseases because it is the most common febrile disease in children. KD and pneumonia show similar fever patterns before the occurrence of respiratory symptoms in pneumonia. 
Despite the small training dataset, the results in our study demonstrated that the deep learning algorithms show excellent performance for the identification of the KD. 
Nevertheless, as the performance of a deep learning algorithm depends on the quantity of training data~\cite{sun2017revisiting, rolnick2017deep}, the deep learning algorithm for KD diagnosis should be extended.

Figure 3 and Figure 4 show the parts of the echocardiographic images that are considered important by the deep learning algorithm to distinguish between KD and non-KD. 
These figures indicate that the explainable deep learning algorithms identified KD by using the features of the coronary arteries. This is comparable to how pediatric cardiologists diagnose and differentiate the diseases. 
Clinical reports have mentioned that coronary artery imaging could be key to the appropriate diagnosis of KD, particularly incomplete KD~\cite{kim2017epidemiology}. 
Through this analysis, our results revealed that deep learning algorithms can identify KD among KD and non-KD, as cardiologists do, which suggests that deep learning algorithms could be applied in a clinical setting to recognize incomplete KD among various acute febrile diseases in children.

Our experimental results showed the potential of explainable deep learning algorithms to distinguish KD from acute febrile diseases. Among the tested algorithms, SE-ResNext50 showed the highest accuracy, specificity, and precision in the classification of KD and non-KD, whereas SE-ResNet50 showed the highest sensitivity. 
However, as shown in the precision-recall-curves, SE-ResNext50 showed the best performance among the algorithms in distinguishing between KD and non-KD. In particular, SE-ResNext50 yielded a sensitivity of 82.64\% and a specificity of 58.12\%. 
This performance of SE-ResNext50 is comparable to that of an experienced cardiologist (sensitivity of 85\% and specificity of 70\%)~\cite{na2019utilization}. 

Thus, these results indicate that explainable deep learning algorithms might be used to diagnose KD at a general hospital without a KD expert. 
In Korea, an experienced pediatric cardiologist is not always available at each hospital, owing to a lack of human resources. Nevertheless, timely diagnosis of KD is essential for proper treatment, to prevent poor outcomes of coronary artery lesions.  

In the global COVID-19 pandemic in particular, there is a risk that KD might be misdiagnosed, as WHO stated that COVID-19 has a similar febrile clinical presentation as KD~\cite{jones2020covid, viner2020kawasaki}. 
Therefore, now more than ever, it is important to distinguish KD from other febrile diseases in children; this may be possible by using an explainable deep learning algorithm. 

Previous studies have analyzed echocardiographic images using deep learning to perform classification of myocardial disease~\cite{zhang2018fully}, detect hypertrophic cardiomyopathy, cardiac amyloid, and pulmonary arterial hypertension~\cite{vidal2021utility}, and evaluate chamber segmentation~\cite{leclerc2019deep} and wall motion abnormalities~\cite{sanchez2018machine, omar2018quantification}. 
However, there has been no study to date on diagnosis of incomplete KD by echocardiographic images of coronary artery lesions using deep learning, as we have done here. This study indicates that explainable deep learning has potential to diagnose KD among acute febrile diseases.
This study had some limitations. 
First, we trained the deep learning algorithms with a small amount of data. 
Second, the experiment was conducted using only pneumonia as a non-KD acute febrile disease. 
To overcome the first and second limitations, more data need to be collected on incomplete KD and other acute febrile diseases, which would be considered as non-KD conditions. 
Training algorithms using such data will provide a KD detection algorithm with higher capability. Our work forms the basis for such future studies.

\section{Conclusions}\label{conc}
We have shown the feasibility of using an explainable deep learning approach for detection of KD based on echocardiography images. 
The AUPRCs of the deep learning algorithms, including VGG19, Xception, ResNet50, ResNext50, SE-ResNet50, and SE-ResNext50, were found to be 0.643, 0.752, 0.822, 0.830, 0.853, and 0.864, respectively, for discrimination between KD and non-KD. 
In particular, the SE-ResNext50 offered the best performance among the deep learning algorithms with an accuracy of 72.88\% and an AUPRC of 0.864. 
The explainable deep learning algorithms highlighted salient features of coronary arteries, similar to how an experienced pediatric cardiologist would examine coronary artery regions for the detection of KD. 
Although the specificity of deep learning algorithms was still lower than that of highly experienced clinicians for the discrimination between incomplete KD and non-KD, the deep learning algorithms used in this study were promising in terms of sensitivity and precision. 
Therefore, deep learning algorithms may assist clinicians in reducing the probability of misdiagnosing KD in clinical practice. 
The abilities of deep learning algorithms should be further developed to be comparable to the performance of highly experienced clinicians in order to translate this approach to application in the clinic.

\section*{Statements of ethical approval}
The study was approved by the Ethics Committee of Severance Hospital.
All the participants provided their written informed consent to participate in this study.

\section*{Declaration of Competing Interest}
There are no conflicts of interest to disclose for publication of this paper.

\section*{Acknowledgements}
This study was supported by a new faculty research seed money grant of Yonsei University College of Medicine for 2020 (2020-32-0035).

\printcredits

{
\bibliographystyle{model1-num-names}
\bibliography{cas-dc-template.bib}
}

\end{document}